\documentclass[pre,aps,showpacs,nofootinbib,twocolumn]{revtex4}
\usepackage{amssymb}

\usepackage{graphicx}

\input{tcilatex}

\begin{document}

\author{Miodrag L. Kuli\'{c}}
\title{Can the bosonic coupling constant
be extracted \\from the ARPES scattering rate in cuprate superconductors?}

\address{Johann Wolfgang Goethe-University, Institute for
Theoretical Physics, \\P.O.Box 111932, 60054 Frankfurt/Main, Germany}

\begin{abstract}
The recent ARPES results for the imaginary part of the self-energy $%
\Sigma ^{\prime \prime }(\omega ,T)$ obtained on a number of HTSC bismuthates 
\cite{Kordyuk} are analyzed. By accepting the ''Fermi-Bose'' \textit{%
division-procedure} of $\Sigma ^{\prime \prime }(\omega ,T)$ into
the Fermi-liquid and bosonic parts - which is proposed in
\cite{Kordyuk}, one obtains very small bosonic coupling constant
$\lambda_{B, Im} <0.2$. 
If this procedure would be correct then the standard Eliashberg
theory makes any bosonic mechanism of pairing irrelevant!
As a consequence we are
\textit{confronted with a trilemma}: (1) to abandon the
``Fermi-Bose'' division-procedure \cite{Kordyuk}; (2) to abandon the
Eliashberg theory; (3) to abandon the interpretation of ARPES data
within the three-step model, where the ARPES intensity is
proportional to the quasiparticle spectral function
$A(\mathbf{k},\omega )$. However, since the bosonic coupling constant
extracted from the ARPES nodal kink at 70 meV \cite{Lanzara},
which measures the real part of the self-energy $\Sigma
^{\prime}(\omega ,T)$, is much larger than the one extracted from
the ARPES line-width ($\lambda_{B,Im}\ll \lambda_{B,Re}>1$) this
means that the ``Fermi-Bose'' division procedure done in
\cite{Kordyuk} is ambiguous.

\end{abstract}

\date{\today}

\maketitle

Recently, very interesting results were reported on ARPES
measurements in a number of HTSC compounds \cite{Kordyuk}, such as
the superstructure free $Bi_{2-x}Pb_{x}Sr_{2}CaCu_{2}O_{8+\delta
}$ ($Bi(Pb)-2212$), $Bi_{2}Sr_{2}CaCu_{2}O_{8+\delta }$
($Bi-2212$) and $Bi_{2}Sr_{2-x}La_{x}Cu_{2}O_{8+\delta }$
($Bi-2201$). By measuring the width ($\Delta k_{FW}(\omega )$) of
momentum distribution curves (MDCs) in the \textit{nodal
direction} the authors of \cite{Kordyuk} were able to extract the
imaginary part $\Sigma ^{\prime \prime }(\omega ,T)$ of the
quasiparticle self-energy $\Sigma (\omega ,T)(=\Sigma ^{\prime
}(\omega ,T)+i\Sigma ^{\prime \prime }(\omega ,T))$ by using the
relation $\Sigma ^{\prime \prime }(\omega ,T)\approx v_{F}\Delta
k_{FW}(\omega )/2$ with $\hbar v_{F}=4$ $eV\AA$ \cite{Kordyuk}.

Let us discuss the results for $\Sigma^{\prime \prime} (\omega)$ in
the \textit{nodal direction} obtained in \cite{Kordyuk} which are presented in 
Fig.1 (which is also
Fig.1 in Ref. \cite{Kordyuk}). In \cite {Kordyuk} the experimental results are
analyzed by  assuming that the ``Fermi-Bose''
division of $\Sigma ^{\prime \prime }(\omega ,T)$ holds, i.e. there is a Fermi-part due
to the Landau-Fermi liquid $\Sigma _{F}^{\prime \prime }(\omega
,T)$ and the Bose-part due to the scattering via the boson channel
$\Sigma _{B}^{\prime \prime }(\omega ,T)$ \cite{Kordyuk}. 
We show below that experiments in Ref. \cite{Kordyuk} give evidence for the significant contribution
of the impurity scattering to $\Sigma (\omega,T)$, which we also take into account. In
that case the self-energy is given by

\begin{equation}
\Sigma ^{\prime \prime }(\omega ,T)=\Sigma _{F}^{\prime \prime
}(\omega ,T)+\Sigma _{B}^{\prime \prime }(\omega ,T)+\Sigma
_{imp}^{\prime \prime }(\omega ,T).  \label{Eq1}
\end{equation}

The Fermi part is given approximately by

\begin{equation}
\Sigma _{F}^{\prime \prime }(\omega ,T)\approx A_{\omega }\omega
^{2}+A_{T}\pi ^{2}T^{2},  \label{Eq2}
\end{equation}
where it is expected that like in the isotropic Landau-Fermi
liquid one has
\begin{equation}
A_{\omega }\approx A_{T}.  \label{Eq3}
\end{equation}

In the inset of Fig.1a the authors in \cite{Kordyuk} determine
$\Sigma_{F} (\omega,T)$ by fitting the data in the
highly overdoped sample ($OD69$ with $T_{c}=69$ $K$) at $T=130$ $K$ \ by $%
Eq.(1)$. Unfortunately, the authors in \cite{Kordyuk} do not report
the value for $A_{\omega }$. Let us determine $A_{\omega }=[\Sigma
^{\prime \prime }(\omega ,T)-\Sigma ^{\prime \prime }(\omega
=0,T)]/\omega ^{2}$ from the data in Fig.1a, i.e. from the data
for $OD69$ at $T=130$ $K$. From the \textit{inset} in Fig.1a one
has$\ \Sigma _{\exp }^{\prime \prime }(\omega =0,130 K)\approx 0.06$ $eV$
and $\Sigma ^{\prime \prime }(\omega =0.3,130 K)\approx 0.25 $ $eV$ \ what
gives a reasonable value $A_{\omega }\approx 2/eV$. 
$\omega$ is given in eV.

\begin{figure}[tbp]
\includegraphics*[width=8cm]{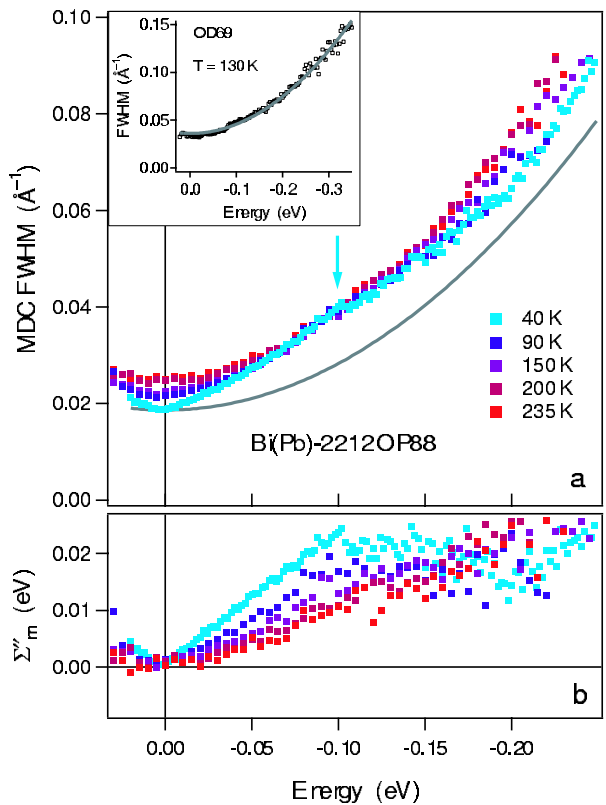}
\caption{T- and $\protect\omega$-dependence of $\Sigma ^{\prime
\prime}$ for the nodal quasiparticles in optimally doped
Bi(Pb)-2212. (\textbf{a}) - the full width at half maximum of the
ARPES intensity. The gray solid line is the Fermi liquid parabola
obtained by fitting the data for highly overdoped sample (OD69) at
130 K (see inset). (\textbf{b}) - the bosonic part $\Sigma
^{\prime \prime} _{B}$ for various T.From \protect\cite{Kordyuk}}
\label{40meVFig}
\end{figure}

Since the authors of \cite{Kordyuk} do not estimate the
contribution of impurities $\Sigma _{imp}^{\prime \prime }(\omega
,T)$ let us do it here. The latter contribution is appreciable
since from the inset in Fig.1 one has that $\Sigma _{imp}^{\prime
\prime }(\omega =0,T)(=\Sigma _{\exp }^{\prime \prime
}(0,T)-\Sigma _{F}^{\prime \prime }(0,T)-\Sigma _{B}^{\prime
\prime }(0,T))$ is large fraction of $\Sigma _{\exp }^{\prime
\prime }(0,T)$. The term $\Sigma _{imp}^{\prime
\prime }(\omega =0,T)$ can be extracted (in a semi-quantitative
way) by considering the experimental results for $\Sigma ^{\prime
\prime }(\omega =0,T)(\gtrsim \Sigma _{imp}^{\prime \prime }(0,0))$ at
$T\ll T_{c}$ . By taking the data from Figs.(1-3) in \cite
{Kordyuk} for $T\ll T_{c}$, we conclude that for a number of systems with $%
T_{c}=(60-90)$ K one has
\begin{equation}
\Sigma _{imp}^{\prime \prime }(\omega =0,T\ll T_{c})\approx (0.02-0.03)eV.
\label{Eq4}
\end{equation}

Let us analyze the effect of impurities on T$_{c}$. Before doing
this, we stress that in the systems which are studied in
\cite{Kordyuk} d-wave pairing is realized. If one assumes that the
standard Eliashberg theory holds and that $\Sigma _{imp}^{\prime
\prime }(0,0)$ is momentum independent, i.e. it contains the
s-wave scattering channel only $\Sigma _{imp}^{\prime \prime
}(0,0)\approx \Sigma _{imp,s}^{\prime \prime }(0,0)$, then this
isotropic impurity scattering is strongly pair-breaking for d-wave
pairing. Since in T$_{c}$ is smaller
than the bare T$_{c,0}$ (for the clean system), then the
interesting question is how big is T$_{c,0}$? The theory
\cite{KulicReview} predicts the following formula for T$_{c}$ in
the case when there is the \textit{s-wave impurity channel} only

\begin{equation}
\ln \frac{T_{c}}{T_{c0}}=\psi (\frac{1}{2})-\psi
(\frac{1}{2}+\rho^{s} _{pb}), \label{Eq5}
\end{equation}
where the pair-breaking parameter $\rho^{s} _{pb}=\mid \Sigma
_{imp,s}^{\prime \prime }\mid /2\pi T_{c}$. By taking the values
for $\Sigma _{imp,s}^{\prime \prime}$ from Eq.(3) one obtains
$T_{c0}^{OD}=(115-150)$ $K$ with $T_{c}^{OD}=69$ $K$ for the
\textit{overdoped} SC, and for the \textit{optimally doped}
$T_{c0}^{OP}\approx (130-160)$ $K$ with $T_{c}^{OP}\approx 90$
$K$.
However, this analysis might be
inappropriate for HTSC, which are strongly correlated materials and which show
surprising robustness of d-wave pairing in the presence of
non-magnetic impurities when the dependence $T_{c}(\rho_{imp})$ is
studied. Here, $\rho_{imp}$ is the residual resistivity - see more
in Ref.\cite{KulicReview}. The theory of the impurity scattering
in strongly correlated systems done in \cite{KulOud}, and which is
based on the theory of strong correlations \cite{KulZey}, shows
the existence of the forward scattering peak in the scattering
amplitude. The latter gives rise to the pronounced 
\textit{d-wave impurity
scattering channel} $\Sigma _{imp,d}^{\prime \prime }$, thus
lowering the impurity pair-breaking effects since in that case
$\rho _{pb}=\mid \Sigma _{imp,s}^{\prime \prime }-\Sigma
_{imp,d}^{\prime \prime }\mid /2\pi T_{c}$. For a further analysis
of the impurity effect on $T_{c}$ the experimental data for
$T_{c}(\rho_{imp})$ are necessary.

The large values of $T_{c}$ (and $T_{c0}$) need also a large bosonic coupling
constant $\lambda _{B}\approx 2$ in the Eliashberg theory. In that respect
one can rise an important question - \textit{ how large is  the
bosonic coupling constant $\lambda _{B}$ extracted from  the ARPES line-width
measurements in \cite{Kordyuk}}? In absence
of a reliable microscopic theory for HTSC oxides one can proceed
by using a phenomenological approach to analyze the ARPES
data. From Fig.1b (also Fig.1b in \cite{Kordyuk}) we see that
the bosonic part of $\Sigma _{B}^{\prime \prime }(\omega )$ is
linear in $\omega $ at low $T$, similarly as in the ''marginal''
Fermi liquid where $\Sigma _{B}^{\prime
\prime }(\omega )\approx (\pi /2)\lambda _{B,Im}\omega $. From the values of $%
\Sigma _{B}^{\prime \prime }(\omega )$ at energies $\omega
>0.05eV$ - where the self-energy is weakly affected by
superconductivity, which we extract from the top curve in Fig.1b
for the optimally doped SC with $T_{c}=88$ $K$ and at $T=40$ $K$, one
obtains a conservative value $\lambda _{B,Im}< 0.2$! Note,  that
the curve at $T=90$ K (slightly above $T_{c}$!) in Fig.1b of Ref. \cite{Kordyuk} - below the top one,
gives even smaller $\lambda_{B,Im}$! Such a small
(bosonic) coupling constant ($\lambda _{B,Im}<0.2$) gives
very small $T_{c}(\ll 100$ $K)$ already for s-wave pairing.
in the standard Eliashberg theory. This means that if the ``Fermi-Bose'' division in
\cite{Kordyuk} would be appropriate than \textit{all
bosonic mechanisms of pairing (EPI, SFI,etc.) would be ineffective
and irrelevant in cuprate superconductors}!

As a consequence we are \textit{confronted with
a trilemma}: (\textbf{1}) to abandon the ``Fermi-Bose''
division-procedure from \cite{Kordyuk}; (\textbf{2}) to abandon the
Eliashberg theory; (\textbf{3}) to abandon the interpretation of
ARPES data within the three-step model, where the ARPES intensity
is proportional to the quasiparticle spectral function
$A(\mathbf{k},\omega )=-\Im G(\mathbf{k},\omega )/\pi $? It seems
that the case (\textbf{1}) is most probable. The argument for this
claim is based on the ARPES measurements of the real part of the
self-energy $\Sigma ^{\prime }(\omega )$. The latter
\cite{Lanzara}, \cite{40meV}, \cite{ShenReview} show \textit{kink}
in the nodal quasiparticle energy at the phonon energies $\omega
\approx 60-70$ $meV$, which gives the coupling constant
$\lambda_{B,Re}=\mid \partial \Sigma^{\prime}/\partial \omega \mid
>1$. The latter coupling is most probably due to the
pronounced \textit{electron-phonon interaction} in HTSC
\cite{KulicReview}. The above analysis shows that
$\lambda_{B,Im}\ll \lambda_{B,Re}$ thus questioning the
``Fermi-Boson'' division procedure done in Ref.\cite{Kordyuk}, which
gives $\lambda_{B,Im}< 0.2$. However, the cases
(\textbf{2})-(\textbf{3}) might interfere too.

In fact, if (i) the (bosonic-like) spin-fluctuation
scattering would be the dominant one - as it is claimed in
\cite{Kordyuk}, and (ii) if the ``Fermi-Bose'' division of
Ref.\cite{Kordyuk} holds, then the ARPES results in
\cite{Kordyuk} tell us that (because $\lambda_{B,Im}\ll 1$) the
spin-fluctuation scattering mechanism is irrelevant
for pairing in cuprate superconductors.
We stress that, there are other reliable arguments against the spin-fluctuation pairing mechanism and 
which are in favor of the
electron-phonon interaction Refs.\cite{KulicReview}, \cite{MaksimovRev}, \cite{KulDolgARPES}.

Finally, we stress, that the small value of the bosonic coupling constant $\lambda_{B,Im}$,
which is extracted from $\Sigma_{ARPES}^{\prime \prime}$ for $\omega >0.05$ $eV$, is common to
all ARPES measurements \cite{ShenReview}. For instance, in the very recent ARPES measurements of the scattering rate 
in optimally and highly overdoped Bi2212 and Bi2201 compounds \cite{Kaminski} it was
found that  $\Sigma_{ARPES}^{\prime \prime}(\textbf{k},\omega)=a_{\textbf{k}}+b_{\textbf{k}}\omega$ 
with $a_{\textbf{k}}$ strongly momentum 
dependent while $b_{\textbf{k}}\approx const=b$ is isotropic. By taking again $v_{F}\approx 4$ $eV\AA$ we obtain
$b\approx 0.4$ and $\lambda_{B,Im} \approx 0.3$. It is hardly to belive that such a small $\lambda_{B,Im}$
can give $T_{c}\approx 100$ $K$ in the Eliashberg theory. 
Therefore, the analysis of the ARPES scattering rate solely by the marginal 
Fermi liquid phenomenology is inadequate and the electron-phonon interaction
must be inevitable taken into account. More on that see in
Refs. \cite{KulDolgARPES}, \cite{Greco}.

\end{document}